\documentclass[12pt]{article}
\usepackage{epsf}
\newcommand{\be}{\begin{equation}}
\newcommand{\ee}{\end{equation}}
\begin{document}

\title{Absence of higher order corrections to the non-Abelian
Chern-Simons coefficient}

\author{F. T. Brandt$^\dagger$, Ashok Das$^\ddagger$ and J. Frenkel$^\dagger$ 
\\ \\
$^\dagger$Instituto de F\'{\i}sica,
Universidade de S\~ao Paulo\\
S\~ao Paulo, SP 05315-970, BRAZIL\\
$^\ddagger$Department of Physics and Astronomy\\
University of Rochester\\
Rochester, NY 14627-0171, USA}

\maketitle

\begin{abstract}

We extend the Coleman-Hill analysis to non-Abelian Chern-Simons
theories containing a tree level topological mass term. We show, 
in the case of a pure Yang-Mills-Chern-Simons theory, 
that there are no corrections to the coefficient of the
Chern-Simons term beyond one loop in the axial gauge. 
Our arguments use constraints coming only from {\em small gauge}
Ward identities as well as the analyticity of the amplitudes, much like
the proof in the Abelian case. Some implications of this result are
also discussed.

\end{abstract}
\vfill\eject

In $2+1$ dimensional QED, with or without a tree level Chern-Simons
(CS) term \cite{1,2}, Coleman and Hill \cite{3}
have proved that the coefficient of the CS term (tree
level or induced) does not receive any quantum correction beyond one
loop at zero temperature. The proof is essentially based on two
key assumptions: i) the Abelian Ward identity and, ii) the analyticity of the
amplitudes in the energy-momentum variables. The Coleman-Hill result
holds whenever these assumptions are valid, but not otherwise. Thus,
for example, in theories with charged massless particles, infrared
divergences  may invalidate the second assumption
\cite{3a}. Similarly, at finite
temperature, amplitudes are known to be non analytic \cite{4} and,
consequently, the Coleman-Hill theorem is known to be violated in this
case \cite{5}. 
Although the work of Coleman and Hill was an attempt to understand
systematically the explicit calculations by Kao and Suzuki \cite{6,7} 
who showed that the two loop correction to the CS coefficient vanishes,
both in the Abelian
as well as in the non-Abelian theories at zero
temperature, the theorem was formulated only for Abelian theories. In
this letter, we extend the result of Coleman-Hill to non-Abelian
theories and show that, using BRST identities as well as the
analyticity of the amplitudes in a Yang-Mills-Chern-Simons
theory, there is no  correction to the coefficient of the CS term beyond one
loop in the axial gauge (more specifically, in an
arbitrary gauge, this result holds only for the ratio of the CS
mass  and the
square of the coupling constant, as we will explain later). It is
worth remarking here that, in a recent paper
\cite{8}, it has been argued, using a generalization of the method of 
holomorphy due to Seiberg \cite{9}, that
in a Yang-Mills theory interacting with matter fields, {\em
without a  tree
level CS term}, there is no higher loop renormalization of the induced CS
coefficient. Our result, for the case with a tree level CS term, is not
covered  by this analysis (as the authors of ref. \cite{8}
specifically point out) and, in fact, this case may be
physically more meaningful. This is because, in the absence of a tree 
level CS term,
infrared divergences in the $2+1$ dimensional theory are so severe
that a loop expansion of the theory may not exist \cite{2,10}.
In such a  case, general formal arguments may be invalidated by 
the infrared divergences of the perturbation theory.

Let us consider the theory described by the Lagrangian density
\cite{2,11,12,13,14}
\begin{equation}
{\cal L}_{\rm inv} = {1\over 2} {\rm tr}\, F_{\mu\nu}F^{\mu\nu} - m\,{\rm
tr}\,\epsilon^{\mu\nu\lambda} A_{\mu}(\partial_{\nu}A_{\lambda} +
{2\,g\over 3}A_{\nu}A_{\lambda})\label{1}
\end{equation}
where we have chosen, for simplicity, the CS mass $m$ to be positive.
The gauge field belongs to a matrix representation of $SU(N)$,
\[
A_{\mu} = A_{\mu}^{a}T^{a}
\]
with the generators of the group assumed to have the normalization
\[
{\rm tr}\, T^{a}T^{b} = -{1\over 2} \delta^{ab}
\]
and
\[
F_{\mu\nu} = \partial_{\mu}A_{\nu} - \partial_{\nu}A_{\mu} + g
[A_{\mu},A_{\nu}].
\]
This is a self-interacting theory and one can, of course, add to it
interacting matter fields. However, we would restrict ourselves,
for simplicity, to the theory described by Eq. (\ref{1}).

Let us briefly comment on some of the essential features of this
theory. First, it is known that, even with a tree level CS term, the
theory is well behaved only in a select class of infrared safe
gauges. In such gauges, the renormalized propagators and
vertices are well defined and
computable, at zero momentum, as  a power series in
${g^2\over 4\pi m}$. 
The infrared safe gauges are linear, homogeneous gauges (with the gauge fixing
parameter $\xi = 0$) and include the Landau gauge as well as the
axial gauges. Second, while in the Abelian theory, the CS coefficient
is a gauge independent quantity and is related to the physically meaningful
statistics factor, in a non-Abelian theory, the CS coefficient is, in
general, gauge  dependent. On the other hand, the renormalization of
${4\pi m\over g^{2}}$ was already calculated earlier 
to one loop order in the Landau gauge \cite{12} and we have verified
that, in all the infrared safe gauges, up to one loop order in this theory,
\begin{equation}
\left({4\pi m\over g^{2}}\right)_{\rm ren} = Z_{m}\left({Z_{3}\over
Z_{1}}\right)^{2}\left({4\pi m\over g^{2}}\right) = {4\pi m\over
g^{2}} + N\label{2}
\end{equation}
where $Z_{3}$ and $Z_{1}$ are the wave function and the vertex
renormalization constants for the gluon, while $Z_{m}$ represents the
renormalization of the CS coefficient. Here, $N$ is the color factor
of $SU(N)$ and this calculation suggests
that this ratio is a physical quantity (it is also this ratio
that needs to be
quantized for {\em large gauge} invariance). Indirectly, we know this
to be true from the fact that, in the leading order in ${1\over m}$
expansion,  i) it is this ratio which determines the
dimensionality of the CS Hilbert space \cite{15} and, ii) this ratio
is related to the coefficient of the WZWN action which represents the
central extension of the corresponding current algebra
\cite{15,16,17}. This also
gives a possible meaning to the one loop result of Eq. (\ref{2}), by
relating it to the product of spin and the dual Coxeter number of the
group \cite{16,17}.

To prove our result, let us choose the axial gauge
\cite{18}
\begin{equation}
n_{\mu} A^{\mu} = 0\label{3}
\end{equation}
which makes the discussion parallel to the Abelian
case. (In this connection, let us also recall that it is in the axial gauge
that the finiteness of the $N=4$ SUSY Yang-Mills theory was
demonstrated \cite{19}.) 
First of all, we know that ghosts decouple in the axial gauge so
that the wave function as well as the vertex renormalizations for the
ghosts are trivial, namely,
\be\label{4}
\widetilde{Z}_{3} = 1 = \widetilde{Z}_{1}
\ee
In fact, in this gauge, the renormalization of the composite sources
involving ghosts is also trivial. As a result, it follows, from the
BRST identities of the theory, that the wave function as well as the
vertex renormalizations for the gluon field satisfy a simple relation,
namely,
\be\label{5}
Z_{3} =  Z_{1}
\ee
In this sense, the theory behaves like an Abelian theory, although the
non-Abelian interactions make the proof more involved. 

The theory, in the axial gauge (\ref{3}), has to be defined carefully
as the limit
$\xi=0$ of the theory with an arbitrary gauge fixing parameter, 
namely, from the theory in a general axial gauge \cite{18}
(As we have already argued, the theory is infrared safe only in this
limit). 
In a general axial gauge with an arbitrary gauge fixing parameter, the
complete two point function for the gluon has the form
\begin{eqnarray}
\Pi^{\mu\nu,ab}(p) & = &
\delta^{ab}\left[\left(p^{\mu}p^{\nu} -\eta^{\mu\nu}\,p^2\right)
(1+\Pi_{1}(p)) +
i\, m\, \epsilon^{\mu\nu\lambda}p_{\lambda} (1+\Pi_{2}(p))\right.\nonumber\\
 &  & \left. + (p^{\mu} -
{p^{2}n^{\mu}\over (n\cdot p)})(p^{\nu}-{p^{2}n^{\nu}\over (n\cdot
p)})\Pi_{3}(p) - {1\over \xi}\,n^{\mu}n^{\nu}\right]\label{6}
\end{eqnarray}
which shows that the self-energy is transverse to the momentum.
(By definition, the self-energy is the two point function
without the tree level term.) Furthermore, 
we see, from eq. (\ref{6}) that the CS coefficient can be obtained from the
two point function as 
\begin{equation}
\delta^{ab}\,(1+\Pi_{2}(0)) = {1\over
6im}\epsilon_{\mu\nu\lambda}\left.{\partial\over 
\partial p_{\lambda}}\Pi^{\mu\nu,ab}(p)\right|_{p=0}\label{7}
\end{equation}
where $\Pi_{2}(0)$ denotes the induced CS coefficient.
However, such a representation is not very useful from
the point of 
view of  an all order proof. Instead, a graphical representation for the CS
coefficient is much more useful and can be obtained through the BRST
identities.

Since the composite sources involving ghosts do not renormalize in the
axial gauge, the BRST (Ward) identities take a simple form, namely, in
the momentum space, we have (all momenta are incoming)
\begin{eqnarray}
 & & p_{n,\mu_{n}}\Gamma^{\mu_{1}\cdots\mu_{n}}_{a_{1}\cdots
a_{n}}(p_{1},\cdots ,p_{n})\nonumber\\
 & = &  ig\sum_{i=1}^{n-1}
f^{aa_{i}a_{n}}\Gamma^{\mu_{1}\cdots\mu_{n-1}}_{a_{1}\cdots
a_{i-1}a\cdots
a_{n-1}}(p_{1},\cdots ,p_{i-1},p_{i}+p_{n},\cdots ,p_{n-1})\label{8}
\end{eqnarray}
which must hold true for each of the external momenta. Furthermore, assuming
analyticity of the amplitudes in the external momenta, we obtain from
Eq. (\ref{8})
\begin{eqnarray}
& & \Gamma^{\mu_{1}\cdots\mu_{n}}_{a_{1}\cdots a_{n}}(p_{1},\cdots
,p_{n-1},0)\nonumber\\
 & = & ig \sum_{i=1}^{n-1} f^{aa_{i}a_{n}}\,{\partial\over \partial
p_{i,\mu_{n}}}\Gamma^{\mu_{1}\cdots\mu_{n-1}}_{a_{1}\cdots
a_{i-1}a\cdots a_{n-1}}(p_{1},\cdots ,p_{n-1})\label{9}
\end{eqnarray}
Using this, as well as Eq. (\ref{7}), it is easy to see that the CS
coefficient can be identified with
\begin{equation}
f^{abc}\,(1+\Pi_{2}(0)) = {1\over 6m g} \epsilon_{\mu\nu\lambda}
\Gamma^{\mu\nu\lambda}_{abc}(0,0,0) \label{10}
\end{equation}
Namely, in this gauge, the CS coefficient can be related to the three
gluon amplitude 
with all external momenta vanishing (this, in fact, makes it quite clear that
the study of this quantity is meaningful only if there are no infrared
divergences in the theory).

Let us note here, from Eq. (\ref{2}) as well as the renormalization
condition in Eq. (\ref{5}), that although the CS
coefficient of a non-Abelian theory is, in general, gauge dependent,
in the axial gauge, it takes on a physical meaning. This happens
because, in this gauge, the renormalized
physical quantity ${4\pi m\over g^2}$ takes the simple form
\be\label{11}
\left(\frac{4\pi m}{g^2}\right)_{\rm ren}= Z_{m}\left({4\pi m\over
g^{2}}\right) = \left(1+\Pi_2(0)\right)\left(\frac{4\pi m}{g^2}\right),
\ee
so that the CS coefficient itself attains a physical significance.
We have checked this explicitly to one loop order and have shown, 
using Nielsen-like identities \cite{4,20}, that
$\Pi_2(0)$ is, indeed, independent of $n^{\mu}$ to all orders.

\begin{figure}
    \epsfbox{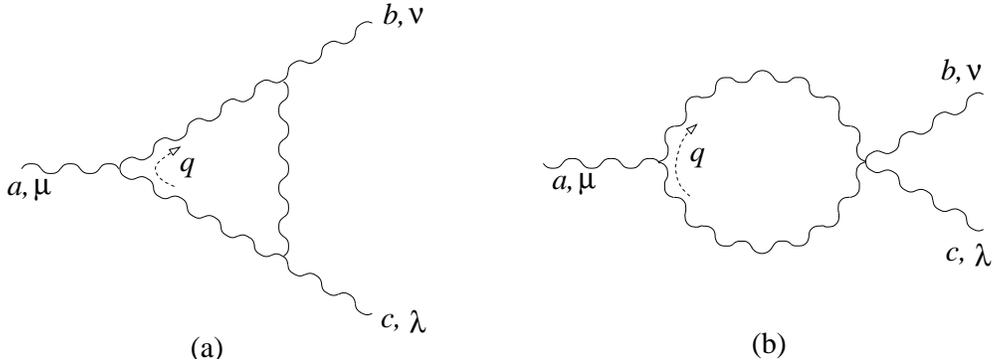}
\caption{One-loop diagrams, with zero external momenta, which
determine the one-loop correction to the  CS coefficient.}
\label{fig1}
  \end{figure}

With the diagrammatic representation of the CS coefficient in the
axial gauge (see Eq. (\ref{10})), let us note that the one loop
correction can be obtained 
from the two diagrams in Fig. 1. The tree level propagator, with a
general gauge fixing parameter, has the form 
\begin{eqnarray}
D^{(0)ab}_{\mu\nu}(q) & = & {\delta^{ab}\over
q^{2}-m^{2} + i\epsilon}
\left[\eta_{\mu\nu} - {q_{\mu}n_{\nu}+q_{\nu}n_{\mu}\over
(n\cdot q)} + {q_{\mu}q_{\nu}n^{2}\over (n\cdot q)^{2}} +
im\epsilon_{\mu\nu\lambda} {n^{\lambda}\over (n\cdot
q)}\right]\nonumber\\
 &  & +\, \xi\,{p_{\mu}p_{\nu}\over (n\cdot p)^{2}}\label{12}
\end{eqnarray}
The propagator in the
axial gauge, is then easily obtained by setting $\xi=0$, when it is
transverse to $n^{\mu}$ (for an alternative method which involves the
use of a Lagrangian multiplier field, see ref. \cite{4}).
With this, as well as the interaction vertices derived from
Eq. (\ref{1}), the evaluation of the diagrams in Fig. 1 is
straightforward, but tedious (with arbitrary $n^{\mu}$) and shows that,
when contracted with $\epsilon_{\mu\nu\lambda}$, the second graph
vanishes, while the first graph gives a non zero
contribution. Explicitly,
\begin{equation}
I^{abc (1)}_{(a)} = f^{abc}\,{g^{3}\over 4\pi}\,6N,\qquad\qquad 
{I}^{abc (1)}_{(b)}= 0\label{13}
\end{equation}
The one-loop correction to the CS coefficient now follows from
Eqs. (\ref{10}) and (\ref{13}) to be
\begin{equation}
\Pi_{2}^{(1)}(0) = {g^{2}\over 4\pi m}\,N\label{14}
\end{equation}
This is gauge independent (independent of $n^{\mu}$) as claimed and,
by the use of relation (\ref{11}), leads immediately to Eq. (\ref{2}).
Eq. (\ref{2}), of course, had been
derived earlier in the Landau gauge \cite{12}, and the present derivation shows
that it holds true in the whole class of infrared safe gauges leading
to the expectation that ${4\pi m\over g^{2}}$ must represent a
physical quantity, as we have argued above.

We would next try to show that the CS coefficient, in the axial gauge,
does not receive any further quantum correction from higher loops. To
this end, we will use the BRST identities in this gauge, namely,
Eq. (\ref{9}) (which, we would like to emphasize, follows from
Eq. (\ref{8}) with the assumption of analyticity).
To simplify our proof, we will use a compact notation, where we treat
the amplitudes as matrices (in the Lorentz and internal symmetry
space). Thus, we define $\Pi$, $D$, $\Gamma^{\lambda}$ and
$\Gamma^{\nu\lambda}$ respectively as the complete two point function, the
propagator, the three point and the four point vertex
functions. In this notation, then, we have
\begin{equation}
\Pi D = -1\label{15}
\end{equation}
and, furthermore, it is straightforward to see, from
Eq. (\ref{9}), that with the external momentum associated with
the index $\lambda$ vanishing, we can write
\begin{eqnarray}
\Gamma^{\lambda} & = & ig\, \partial^{\lambda}\Pi\nonumber\\
{\rm or,}\;\;\; D\Gamma^{\lambda}D & = & ig\, D(\partial^{\lambda}\Pi)D =
ig\, \partial^{\lambda}D\label{16}
\end{eqnarray}
Here and in what follows, $\partial^{\lambda}$ represents the
derivative  with respect to the appropriate momentum and we have
ignored writing out explicitly the internal
symmetry factors for simplicity. (Namely, the internal symmetry
factors simply come out of the integral and are not relevant to our
proof, as will become evident shortly.) There are two classes of
diagrams (shown in Figs 2 and 3) which can contribute to higher order
corrections of the CS coefficient. Using relations
(\ref{15})-(\ref{16}), which  hold to
any order in perturbation theory, it is now
straightforward to show that higher loop corrections (beyond one loop)
to  the CS coefficient, coming from one particular class of diagrams, vanish. 

\begin{figure}[t!]
    \epsfbox{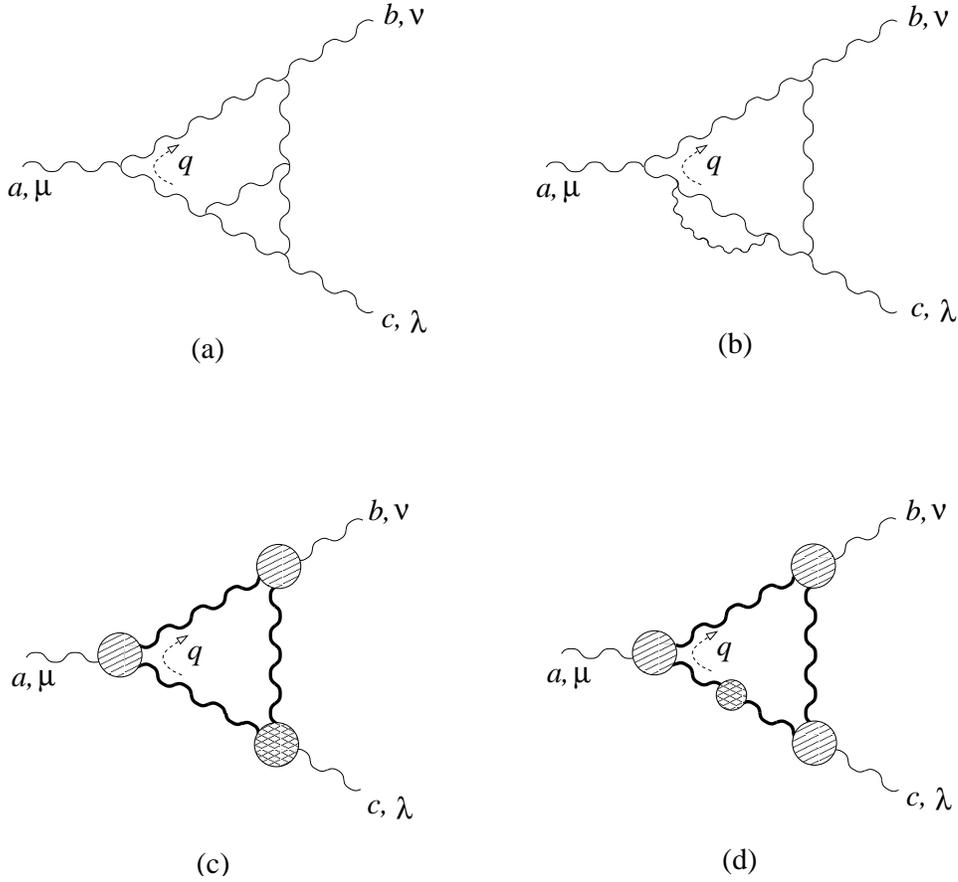}
\caption{Examples of two-loop diagrams
[(a),(b)] which contribute to the three gluon amplitudes [(c),(d)].
Graphs obtained by cyclic permutations of the external
gluons are understood to be included.}
\label{fig2}
\end{figure}

Let us
consider the diagrams in Figs. 2c and 2d, where all external momenta vanish.
Here, the hatched vertices and the bold internal lines represent
respectively the three point vertices and the propagators which
include all the corrections up to $n$-loop order, with
$n=0,1,2,\cdots$. The cross-hatched vertex includes all the
correction up to $(n+1)$-loop order starting from one-loop (namely, it
does not contain the tree level term), while the
cross-hatched loop in the internal propagator stands for the self-energy,
which includes  all the corrections up to  $(n+1)$-loop order. 
We will put an overline on these two
factors just to emphasize this aspect, namely, that they do not
contain the tree level contribution. 
By definition, therefore, the diagrams in Figs. 2c and 2d give
contributions at two loops and higher. Furthermore, from the definition
given above, we can write, with the notation described earlier,
\begin{equation}
\overline{\Gamma}^{\lambda} = ig\,\partial^{\lambda}\overline{\Pi}\label{17a}
\end{equation}
The contributions from these
diagrams would yield a part of the $(n+2)$ loop corrections to the CS
coefficient. Contracting the three
point amplitudes in Figs. 2c and 2d with
$\epsilon_{\mu\nu\lambda}$, we obtain,
\begin{eqnarray}
I^{(n+2)}_{(2c)+(2d)} & = & 
{\rm Tr} \int d^{3}q\,\epsilon_{\mu\nu\lambda}
\left[\left(D \Gamma^{\mu} D \Gamma^{\nu}  D
\left(
\overline{\Gamma}^{\lambda}+\Gamma^{\lambda} D\overline{\Pi}
\right)\right)^{(n+1)}
+ {\rm cyclic} \right]\nonumber \\
 & = & - i\,g^{3}\,{\rm Tr} \int d^{3}q\,\epsilon_{\mu\nu\lambda}
\left[\left(\partial^\mu D \partial^\nu \Pi
\left(
D\partial^\lambda \overline{\Pi} +
\partial^\lambda D\overline{\Pi}\right)
\right)^{(n+1)} + {\rm cyclic}\right]\nonumber \\
 & = &  -i\,g^{3}\,{\rm Tr} \int d^{3}q\,\epsilon_{\mu\nu\lambda}
\left[\partial^\lambda
\left(
\partial^\mu D\partial^\nu\Pi D\overline{\Pi}
\right)^{(n+1)} + {\rm cyclic}\right]= 0\label{17} 
\end{eqnarray}
for all $n=0,1,2,\cdots$, where the superscript, $(n+1)$, stands for
the order  of the terms in the expression. Here, 
\lq\lq Tr'' denotes  trace over the matrix indices in the Lorentz space and
we have used the identities in Eqs. (\ref{15}),(\ref{16}) and
(\ref{17a}) in deriving Eq. (\ref{17}). (There are
also matrix indices associated with the internal symmetry space which
are not traced, but they
are not relevant for our argument as is evident). 
We note that, because of the epsilon tensor, the factor inside the
divergence picks out
only the parity violating terms of the amplitude, which converge
sufficiently rapidly to zero as $q\rightarrow\infty$.
This shows that all the higher loop corrections
(two loop and above), to the CS coefficient, coming from this class of
diagrams vanish.

There is the second class of diagrams, shown in Fig. 3c, which can also
contribute to the CS  coefficient. From
the  identities in Eq. (\ref{9}), we can  express the four
point vertex, with two external momenta vanishing, in terms of the
three point vertex with one external momentum vanishing in a compact
form as (all the momenta are incoming)
\begin{equation}
\Gamma^{\nu\lambda}(0,0,;q,-q)=
ig\left[\frac{\partial}{\partial q_\lambda}
\Gamma^{\nu}(q^\prime-q;q,-q^\prime)\right]_{q^\prime=q} 
\equiv ig\,\partial^{\lambda}\Gamma^{\nu}\label{18}
\end{equation}
where again, we have suppressed the internal indices and we follow the
convention that the momenta associated with the indices $\nu,\lambda$
of the four point vertex as well as that associated with the index
$\nu$ of the three point vertex vanish. Written out
explicitly, the right hand side of Eq. (\ref{18}) would involve two
terms  with different
distributions of the internal indices, but, as we have emphasized
earlier, the internal symmetry factors are not very relevant
to the proof of our result. 

With these, let us look at the 
class of graphs in Fig. 3c, with all external momenta vanishing. 
As opposed  to the diagrams in Fig. 2c and 2d, here all the 
vertices and the propagators include corrections to all orders
(namely, they are the full vertices and propagators of the theory). With the
use of Eqs. (\ref{16}) and (\ref{18}),
the contraction of $\epsilon_{\mu\nu\lambda}$ with the amplitude in
Fig. 3c  yields
\begin{eqnarray}
I_{(3c)} & = & {\rm Tr} \int d^{3}q\,\epsilon_{\mu\nu\lambda}
D\Gamma^{\mu}D\Gamma^{\nu\lambda}\nonumber
= - g^{2}\,{\rm Tr} \int d^{3}q\,\epsilon_{\mu\nu\lambda}\,
\partial^{\mu}D\partial^{\lambda}\Gamma^{\nu} \nonumber \\
& = &  -g^{2}\,{\rm Tr} \int
d^{3}q\,\partial^{\mu}\left(\epsilon_{\mu\nu\lambda}
D\partial^{\lambda}\Gamma^{\nu}\right) =  0\label{19} 
\end{eqnarray}
\begin{figure}[t!]
    \epsfbox{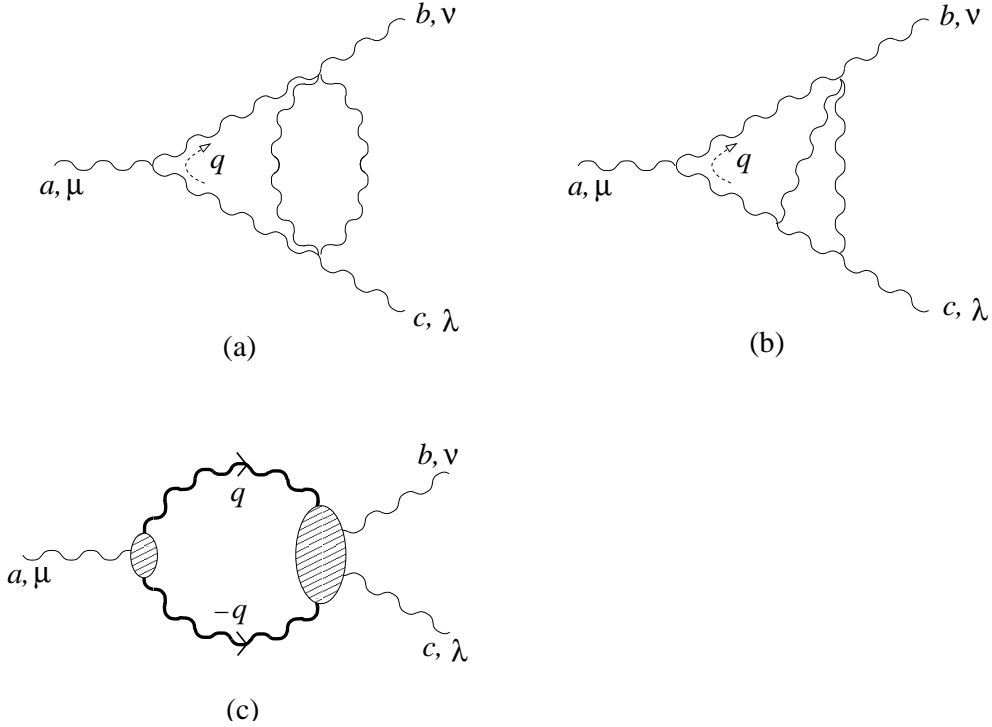}
\caption{Examples of two-loop Feynman diagrams [(a),(b)] which
contribute to the class of graphs (c).}
\label{fig3}
  \end{figure}
Once again, the integrand in Eq. (\ref{19}) is sufficiently convergent
(because it involves only the parity violating parts of the
amplitude) so that the integral vanishes. We have, of course, already
seen explicitly, at the one loop level (in Eq. (\ref{13})),
that this diagram gives a vanishing contribution to the CS
coefficient. Eq. (\ref{19}) shows that this class of diagrams do not
contribute to the CS coefficient at all.

Since these are all the diagrams that can contribute to the higher loop
corrections of the CS coefficient, we have shown that, in a
Yang-Mills-Chern-Simons theory, the CS coefficient, in the axial
gauge, does  not receive any correction beyond one loop
order.  In other
words,  much
like the proof in  the Abelian theory \cite{3}, we have used the
non-Abelian Ward  identities in the
axial gauge, together with the analyticity of the amplitudes in momentum
space, to show that the CS coefficient has no quantum correction
beyond one-loop in this gauge. In theories where these assumptions are
valid, we will expect our proof to hold true. On the other hand, if
either of these assumptions is violated, the proof is expected to
break down, as
would be the case, for example, at finite temperature.

Let us note that, in view of relation (\ref{11}),
this result also means that ${4\pi m\over g^{2}}$ has
no correction beyond one loop in this gauge. On the other hand, as we
have  argued,
this is a physical quantity and, therefore, this result must hold true in any
other infrared safe gauge such as the Landau gauge and, consequently,
Eq. (\ref{2}) must be exact in such a theory in any infrared safe
gauge. We know, however, that, in other non-axial type
gauges, such as the Landau gauge, the wave function and the
coupling constant renormalizations are not related in a simple manner
as in Eq. (\ref{5}). 
Consequently, it follows that, in other
gauges, the CS coefficient itself will receive higher loop
corrections. But these higher loop corrections must be related 
to the wave function and the vertex renormalizations of the
gluon field in such a way that ${4\pi m\over g^{2}}$ has vanishing
contribution beyond one loop.

Such a result has, of course, been expected and predicted. In fact, there is a
plausibility argument for this, based on {\em large gauge} invariance
in the  following way \cite{6,12}. The only
dimensionless ratio in this theory is ${g^{2}\over 4\pi m}$ ($4\pi$ is
a normalization) and can be
used as a perturbative expansion parameter. With this, we can write,
\begin{equation}
\left({4\pi m\over g^{2}}\right)_{\rm ren} = {4\pi m\over g^{2}}
\sum_{n=0}^{\infty} a_{n}(N)\left({g^{2}\over 4\pi
m}\right)^{n}\label{20}
\end{equation}
with $a_{0}(N)=1$ and, as we have seen, $a_{1}(N)=N$. On the other
hand, the invariance of the Chern-Simons
term under {\em large gauge} transformations requires that the ratio
${4\pi m\over g^{2}}$ be quantized, both in the bare
as well as in the renormalized theory (they don't have to be the same
positive integer). Clearly, this is possible for arbitrary 
integers and color factors, only if the series, on the right hand
side of Eq. (\ref{20}), terminates after the second term. Our proof
explicitly verifies that this expectation is, indeed,
justified. However, it is important to recognize that our proof uses
constraints coming only from the behavior under {\em small gauge}
transformations (and, of course, analyticity), much like the proof in
the Abelian case.

To summarize, we have shown, using the BRST identities as well as the
analyticity of amplitudes, that the CS coefficient does not receive
any correction beyond  one loop  in the axial gauge. 
We have verified this behavior by an explicit calculation, which shows
that all the two loop
contributions to the CS coefficient do indeed add up to zero in this gauge.
This allows us to  conclude that the ratio ${4\pi m\over g^{2}}$ is not
renormalized beyond one loop in any infrared safe gauge. For lack of
space, we have only sketched our proof and announced various
results. The details of the calculation with many other aspects of
this problem will be published separately \cite{21}.

We would like to thank Gerald Dunne and Roman Paunov for some useful
discussions. This work was supported in part by U.S. Dept. Energy Grant DE-FG
02-91ER40685, NSF-INT-9602559 as well as by CNPq, Brazil. 

\providecommand{\href}[2]{#2}\begingroup\raggedright
\endgroup

\end{document}